\documentstyle[twocolumn,aps,prl]{revtex}
\begin{document}

\draft

\twocolumn[\hsize\textwidth\columnwidth\hsize\csname @twocolumnfalse\endcsname

\title{Molecular conduction: paradigms and possibilities}

\bigskip

\author{A. W. Ghosh and S. Datta}
\address{ School of Electrical and Computer Engineering, Purdue
University, W. Lafayette, IN 47907}%

\maketitle

\medskip

%\author{P. S. Damle, A. W. Ghosh and S. Datta}
%\address{ School of Electrical and Computer Engineering, Purdue
%University, W. Lafayette, IN 47907}%

\widetext
\begin{abstract}
We discuss the factors that determine the overall shape and magnitude of
the current-voltage (I-V) characteristics of a variety of molecular conductors
sandwiched between two metallic contacts. 
We analyze the individual
influences of the contact geometry, the molecular chemistry, the electrostatics of
the environment, and charging on molecular conduction. 
The theoretical predictions depend sensitively 
on the experimental geometry, as well
as on the theoretical model for the molecule and the contacts. Computing molecular
I-V characteristics will thus require theoretical understanding on several fronts,
in particular, in the {\it{scheme for calculating the molecular energy levels}}, as
well as on the {\it{position of the contact Fermi energy relative to those levels.}}
\end{abstract}
\bigskip

%\pacs{PACS numbers: 85.65.+h, 73.23.-b,31.15.Ar}
%31.15.Ar Ab initio calculations
%81.07.Nb Molecular Nanostructures
%81.07.Lk Nanocontacts
%85.65.+h Molecular electronic devices
%72.10.Bg General formulation of transport theory
%72.20.Dp General theory, scattering mechanisms of conductivity
%73.23.-b Electronic transport in mesoscopic systems
%73.40.Sx Metal-semiconductor-metal structures
%73.63.-b Electronic transport in mesoscopic or nanoscale materials and structures
%2col
%end of wide text
]
\narrowtext
%2col

\section{Molecular conduction: what is the underlying physics?}

Recently several researchers have measured charge transport in
single or small groups of organic molecules connected to metal contacts
\cite{rReed,rChen,rDatta,rCollier,rReichert,rDhirani,rLindsay,rDekker,rMetzger}.
In parallel, there have been theoretical attempts at
understanding molecular conduction, both at the semi-empirical
\cite{rDatta,rEmb,rManoj,rPauls,rJoachim,rHall,rYaliraki,rCuniberti}
and first-principles
\cite{rDamle,rSeminario,rPantelides,rGuo,rPalacios,rPalacios2,rBrandbyge} levels.
Understanding molecular conduction is challenging, since it involves
not just the intrinsic chemistry of the molecule, but extrinsic factors
as well, such as the metal-molecule bonding geometry, contact surface
microstructure and the electrostatics of the environment. The aim of
this article is to discuss the various physical factors that influence
molecular current-voltage (I-V) characteristics, and our attempts to
model them both qualitatively and quantitatively.

A typical two-terminal molecular I-V looks like Fig.~1, often with a clear
conductance gap \cite{rReed}. How does one understand such an I-V?  The
first step is to draw an energy-level diagram, as in Fig.~2. An isolated
molecule has a discrete set of energy levels, with a highest occupied
(HOMO) and a lowest unoccupied molecular orbital (LUMO), separated by a
HOMO-LUMO gap (HLG). On connecting the molecule to metallic contacts,
two changes happen: (i) the discrete molecular levels broaden into a
quasicontinuum density of states (DOS) due to hybridization with the
metal wave functions. Often the DOS retains a distinct peak
structure, in which
case it is still useful to think in terms of broadened molecular energy
``levels''; (ii) the difference in work functions between the molecule and
the metal causes charge transfer and band alignment between the two materials. The molecule
equilibrates with the contact with an overall chemical potential set
by the metal Fermi energy $E_F$, typically lying inside the HLG. Under
an applied bias the molecule tries to equilibrate simultaneously with
both contacts with bias-separated chemical potentials $\mu_{1,2}$, and is
thereby driven strongly out of equilibrium.  As long as the bias is small
and $\mu_{1,2}$ lie in the HLG, the HOMO levels stay filled and LUMOs are
empty and there is no current. However, when the bias is large enough that
either $\mu_1$ or $\mu_2$ crosses a molecular level $E_{\rm{MOL}}$,
that level is filled (reduced) by one contact and emptied (oxidized)
by the other (Fig. 2a) and therefore starts conducting current \cite{rBirge}. For
opposite bias, the {\it{same}} level starts conducting when crossed
by the other contact chemical potential (Fig. 2b). The net result is that for
a spatially symmetric molecule with symmetrically coupled contacts
the {\it{total}} conductance gap is given by $\sim 4\left(E_F -
E_{\rm{MOL}}\right)$.  {\it{Molecular conduction thus depends on both
the intrinsic molecular chemistry through $E_{\rm{MOL}}$ and the contact
microstructure through $E_F$.}}

\begin{figure}
\vspace{3.0in}
%\hskip 0.2cm{\vskip 6cm\special{psfile=GVparams.eps
%\hskip 2cm{\vskip 5.2cm\special{psfile=GV_params.eps
{\includegraphics{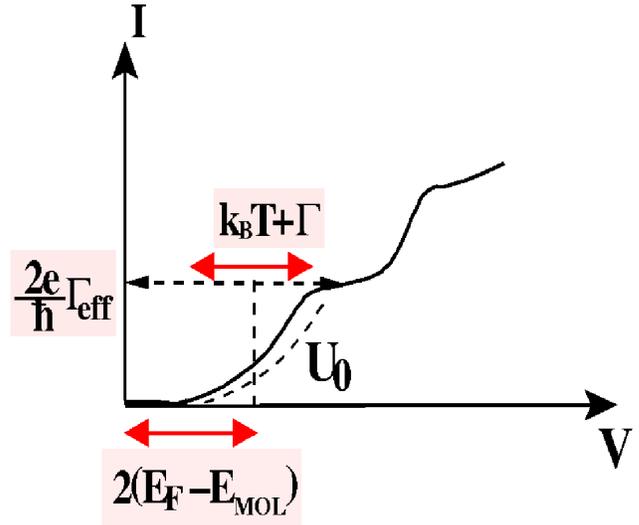}}
\caption{Generic molecular I-V and parameters controlling it. The
current rises when a molecular level $E_{\rm{MOL}}$ is crossed at a bias $V \approx
2\left(E_F - E_{\rm{MOL}}\right)$, where $E_F$ is the contact Fermi energy.
The overall current magnitude is controlled by $\Gamma_{\rm{eff}} 
= \Gamma_1\Gamma_2/[\Gamma_1 + \Gamma_2]$, 
$\Gamma$s being the broadenings of
the molecular levels by hybridization with the contacts. The
current rises over a voltage width set by the thermal broadening $k_BT$ and by
$\Gamma = \Gamma_1 + \Gamma_2$. The current is dragged out further by
the presence of a Coulomb charging energy $U_0$.}
\label{f1}
\end{figure}

The intrinsic chemistry of an isolated molecule can be handled
with sophisticated quantum chemical codes that can be purchased or
even downloaded from the Internet. Given an appropriate basis-set
(for instance, a minimal STO-3G basis) and an appropriate model for
electron-electron interactions (based on first-principles density
functional, Hartree-Fock or semi-empirical H\"uckel methods), such a
code starts with an initial guess density matrix $\rho$ to obtain a
Fock matrix $F$ (Fig. 3a). It then fills up the corresponding energy
eigenstates with a given number $N$ of electrons according to equilibrium
statistical mechanics to reevaluate $\rho$, recalculates $F$ and so
on, until self-consistent convergence. Our molecular system differs
in two ways: (i) it is open, with a varying, fractional occupancy of
electronic levels; (ii) it is trying to equilibrate under bias with
two different contact chemical potentials and is therefore driven
strongly out of equilibrium. To solve this problem, we modify the above
self-consistent scheme, as shown in Fig. 3b.  The initial step, solving
for the Fock matrix $F$, is kept unchanged from the usual prescriptions
in molecular chemistry. In this step, one can use semi-empirical
tight-binding/H\"uckel-based methods \cite{rHoffman}, or exploit the
sophisticated numerical prowess of a standard quantum chemical package
such as GAUSSIAN'98 \cite{rGaussian} to employ a density functional theory
(DFT)-based method for evaluating an $F$ matrix. The $F$ matrix is then
supplemented with self-energy matrices $\Sigma_{1,2}$ describing an open
system connected to the two contacts and involving the detailed contact
microstructure, while the nonequilibrium (transport) part is set up
using the nonequilibrium Green's function (NEGF) formalism \cite{rXue}.
We refer the reader to our past work for details of the NEGF equations and
the calculation of the self-energy matrices \cite{rMorkoc,rCRC,rDamle}. In
this article, we will concentrate mainly on the physical insights.

\begin{figure} \vspace{2.1in} \hskip 1cm{\includegraphics{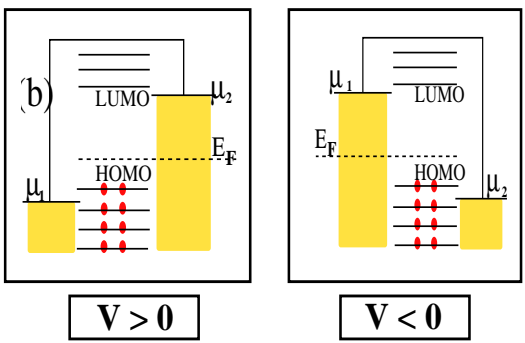}} \caption{Onset of conduction is given by the
voltage where either of the contact chemical potentials $\mu_{1,2}$
crosses the nearest conducting molecular level, HOMO in
this case.} \label{f2} \end{figure}

The coupled DFT-NEGF formulation of molecular electronics is in effect
the generalization of the coupled Poisson-hydrodynamic equations used
extensively in analyzing device transport
\cite{rBaranger,rSelberherr}.  We supplement the Poisson (Hartree) 
term with exchange-correlation corrections that are small in
macroscopic devices but significant in molecules, while the semiclassical
hydrodynamic equation is replaced with a fully
quantum mechanical NEGF formalism \cite{rDattabook}.

The self-consistent formalism described above is completely general,
and can be employed in modeling transport through a wide variety of
physically different systems. For instance, we have used this scheme
to obtain semi-empirical \cite{rDatta,rReedchap} and first-principles
density functional (DFT)-based (LANL2DZ/B3PW91) \cite{rDamle,rDamGhosh}
I-V characteristics of metallic quantum point contacts as well as
semi-conducting aromatic thiol molecules bonded to Au(111) surfaces.
Whether one is dealing with a carbon nanotube, a ballistic MOSFET
\cite{rReedchap,rVenugop}, a spin transistor, a resonant tunneling diode
\cite{rLake} or a molecular wire, the above formalism holds.  One needs
simply to evaluate each of the following quantities: (i) an appropriate
Fock matrix $F$ describing the device; (ii) a contact Fermi energy $E_F$
relative to which the device levels are evaluated and which determines
the electrochemical potentials $\mu_{1,2}$; (iii) a self-consistent
potential $U_{\rm{SCF}}$ describing charging effects and the electrostatic
influence of the environment (this term is incorporated into the effective
$F$ matrix of the device), and (iv) a set of self-energy matrices
$\Sigma_{1,2}$ that describe the coupling of the device with the contact,
the matrices depending on the contact surface Green's function as well
as the device-contact bonding geometry. Additional self-energy matrices
can be introduced to describe scattering, by phonons or polarons for
example. Within the same self-consistency scheme and NEGF prescription,
one can then get qualitatively different I-V characteristics just by
varying $F$, $E_F$, $U_{\rm{SCF}}$ and $\Sigma$.

\begin{figure} \vspace{1.4in} {\includegraphics{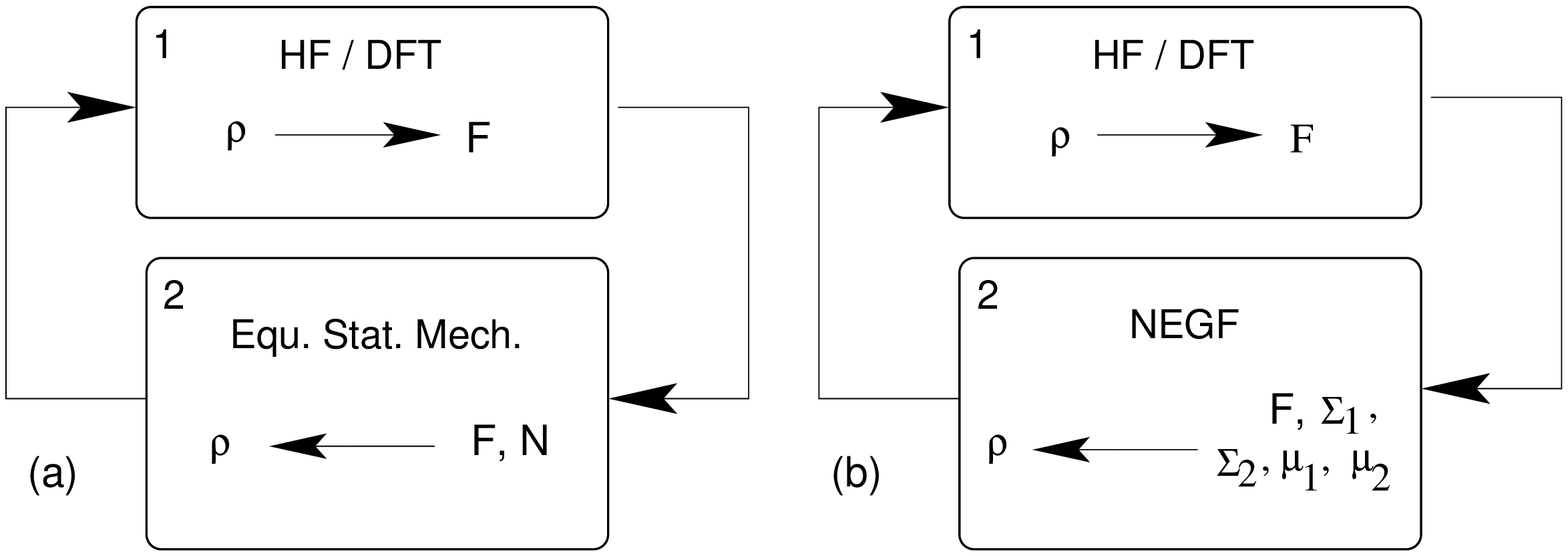}} \vskip 0.5cm \caption{Self-consistency schemes: (a) for an
isolated molecule in equilibrium, one calculates the Fock matrix $F$ starting
with a guess density matrix $\rho$, and fills up the corresponding
levels with $N$ electrons to get back $\rho$; (b) for an open system,
the molecular Fock matrix is supplemented with self-energy matrices
$\Sigma_{1,2}$ describing coupling with the contacts. An applied bias
drives the system out of equilibrium due to two different contact chemical
potentials $\mu_{1,2}$. The step from $F$ to $\rho$ is different from (a),
and is obtained by solving the NEGF equations \protect\cite{rDattabook}.}
\label{f3} \end{figure}

The effect of each of the above parameters on the molecular I-V is
schematically shown in Fig. 1. The conductance gap depends on
$\left(E_F - E_{\rm{MOL}}\right)$, and the maximum current level is
set by the {\it{parallel combination}} $\Gamma_{\rm{eff}} =
\Gamma_1\Gamma_2 /(\Gamma_1 + \Gamma_2)$ of the individual contact
broadenings $\Gamma_{1,2} = i(\Sigma_{1,2} - \Sigma_{1,2}^\dagger)$.
The current rises over a width which depends on the total broadening of
the molecular levels, which in turn depends on (a) the thermal
broadening $k_BT$, (b) the {\it{series combination}} $\Gamma = \Gamma_1
+ \Gamma_2$ of the contact broadenings $\Gamma_{1,2}$ and (c) the
Coulomb charging energy $U_0$ to add an extra charge to the molecule
(the charging energy tends to drag out the conductance peak, so that
for appreciable charging energies, the contact chemical potentials may
not be able to cross a molecular level easily under bias, resulting in
a relatively featureless I-V characteristic). Finally, the potential
profile across the molecule sets the overall voltage division factor
$\eta$ \cite{rDatta} which determines the prefactor in the ratio between the
conductance gap and $E_F - E_{\rm{MOL}}$ ($\eta = 0.5$ and the prefactor
equals 4 for symmetric
coupling, as in Fig. 1). The voltage division factor $\eta$ depends on
the contact geometries and characterizes the Laplace part of the
potential, while the Poisson part describes self-consistent charging
effects, and is characterized by $U_0$.

We summarize below the most challenging and physically relevant questions
in obtaining a molecular I-V characteristic:

\begin{itemize} \item{} {\bf{Where is the contact Fermi energy relative
to the molecular levels?}} $(E_F, E_{\rm{MOL}})$

\item{}  {\bf{What is the broadening due to the contacts?}} $(\Gamma_{1,2})$

\item{}  {\bf{What is the spatial profile of the Laplace potential?}}
$(\eta)$

\item{} {\bf{What is the charging energy?}} $(U_0)$

\end{itemize}

Note that each of the above quantities is in general a complicated
matrix that can be modeled independently using either semi-empirical or
first-principles methods. However, in order to develop a ``feel'' for
how these quantities affect the molecular I-V as in Fig. 1, we 
will try to capture their essence in terms of a
few characteristic scalar parameters, as defined above. We will now
address the influence of each parameter one at a time below.

\section{Molecular conduction: how can we model it?}

\subsection{Where is the contact Fermi energy relative to the molecular
levels?}

This is probably the most challenging problem to sort out. One needs
to start by modeling the quantum chemistry of the molecule. For our
candidate molecule phenyl dithiol (PDT), shown in Fig.~4, it is believed
that the sulphur atoms bond with a Au(111) surface by desorption of
the end hydrogen atoms that are then replaced by a triangle of gold
atoms with sulphur sitting above their centroid \cite{rgeom}. The energy
levels of the isolated PDT molecule compare well with those obtained by
replacing each H-atom by three gold atoms (the gold atoms introduce some
additional localized levels in the HLG).  Replacing the gold cluster with
a self-energy describing metallic gold broadens the molecular levels
into a quasicontinuous spectrum, with the localized levels developing
into metal-induced gap states (MIGS) decaying spatially away from
the contacts into the molecular center.  Now, the energy levels of the
isolated molecule itself depend sensitively on the method of calculation
(a comparison plot is shown in Fig. 4). Different theoretical groups
have adopted different first-principles schemes in their analysis
\cite{rDamle,rSeminario,rPantelides,rGuo}, so the unanswered question
at this point is: {\it{which method is appropriate for calculating the
single-particle energy levels of an open subsystem under bias?}}

\begin{figure} 
\vspace{3.1in} 
%{\special{psfile=PDTlev.ps
%{\special{psfile=HLG.ps 
{\includegraphics{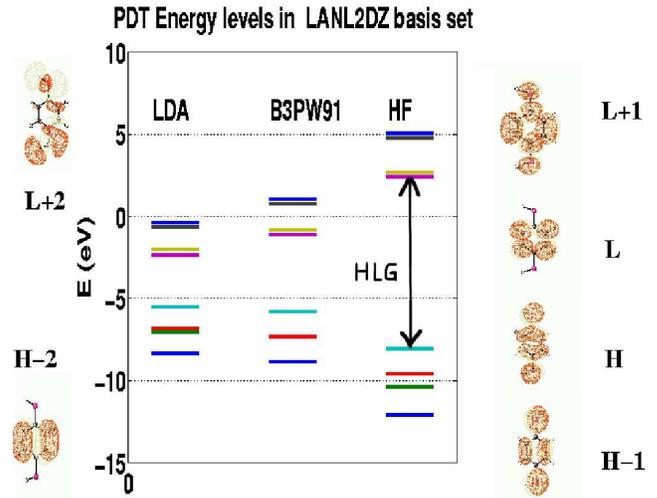}} 
\label{f4} 
\caption{The single-particle molecular
energy levels of PDT vary considerably for density functional-based and
Hartree-Fock methods, giving different conductance gaps. The orbital wave functions
agree for all the different methods and for various choices of basis
sets (symbols H:
HOMO, L: LUMO, HLG: HOMO-LUMO gap, LDA: Local Density Approximation,
B3PW91: 3 parameter Becke exchange and Perdew-Wang 91 correlation, HF:
Hartree-Fock ).} 
\end{figure}

What is consistent among the various methods of calculation, including
semi-empirical (H\"uckel-based) theories, is the overall shape of the
orbital wave functions; for instance, the HOMO is largely sulphur-based
and delocalized while the LUMO is ring-based and localized. This
would seem to suggest a broad HOMO DOS and a sharp LUMO DOS, although
theories don't seem to agree even on this point (see for example Fig.
22 in \cite{rMorkoc} and Fig. 3 in \cite{rPantelides2}). The difference
could arise due to different bonding geometries assumed at the gold-thiol
interface. Thus the intrinsic chemistry of the molecular system needs
to be cleared up, and theoretical agreement reached on the equilibrium
molecular bonding properties before the transport-related issues
can be sorted out.

The position of the contact Fermi energy $E_F$ relative to the molecular
energy levels is also an unsettled issue. The Fermi energy depends
sensitively on the specific model for the contact geometry. Different
models for the contact (for instance DFT-based \cite{rGuo,rDamle},
Bethe lattice \cite{rPalacios}, jellium \cite{rPantelides}) can give
different charge contents and level broadenings of the molecule, as well
as different work functions for the bulk electrodes. This could cause
an appreciable shift in $E_F$, given the rather small gap density of
states in the molecular HLG.  Given that for extensively studied systems
such as metal-semiconductor interfaces the precise location of $E_F$
is still an active topic of research \cite{rEF}, perhaps the best one
can do at this point is to inquire if $E_F$ is closer to the molecular
HOMO or LUMO level, the analogous question for a semiconductor being
whether it is p-type or n-type.

Conceptually the cleanest way to address the equilibrium Fermi energy
problem is by including a few layers of the contact as a cluster in a
``supermolecule", which would act as the device under investigation.
Thereafter the contact is assumed unaltered during conduction, with all
the ``action'' lying in the device sector.  The advantages of including
such a cluster are enormous (for a discussion, see \cite{rDamGhosh}),
such as the automatic inclusion of image charges (the supermolecule is
charge neutral), avoiding uniqueness issues related to partitioning in a
non-orthogonal, non-localized, atomic basis set and the proper treatment
of the surface physics. Ideally, the cluster size should be significantly
larger than the atomic Debye length of the contact material, while for
practical purposes, it is usually limited by computational resources.
$E_F$ is usually set by the HOMO of the (large) supermolecule,
while the molecular levels can be identified by either plotting the
wavefunctions or by computing the local density of states (LDOS) on the
molecule. To employ this scheme to sort out the molecular chemistry and
energy level structure, it is essential that the contact cluster and the
molecule be calculated using the {\it{same}} scheme (DFT/tight-binding,
etc).  Attempts at performing such a computation at the semi-empirical
\cite{rEmb} and DFT \cite{rWeb} levels have yielded a Fermi energy quite
close to the HOMO level for PDT-Au(111) heterostructures.

The experimental situation is somewhat unclear. The conductance gap for
PDT itself is different for different experimental geometries.  The gap
is around 3 volts for breakjunction measurements by Reed {\it{et al.}}
\cite{rReed} and around 4 volts for STM measurements by Hong {\it{et al.}}
\cite{rHong}, while corresponding breakjunction measurements by David
Janes' group at Purdue indicate featureless I-V characteristics with
no discernible conductance gap \cite{rJanes}. Since the gap depends on
$E_F$ which is likely to be different for the two experimental contact
geometries, and could further be compromised by the presence of charging
in the system, such a difference is not surprising. It is also not
clear whether the conduction is through a single molecule bridging the
junctions, or a series combination of molecules attached separately to
the two junctions \cite{rKircz}. It seems sensible therefore to treat
$E_F$ as a fitting parameter, in the absence of precise characterization
of the contact surfaces and molecular geometry. Alternatively, one could
dictate the position of $E_F$ relative to the levels, guided by separate
equilibrium cluster calculations, as discussed earlier.

Experiments incorporating a third terminal (gate) can help clarify some of
these issues appreciably. For instance, a positive gate voltage lowers the
molecular levels so that the Fermi energy approaches the LUMO and moves
away from the HOMO. For a purely electrostatic gate control mechanism,
a measured decrease in conductance will suggest HOMO (p-type) conduction,
while the reverse result indicates LUMO (n-type) conduction.

\medskip \noindent {\bf{Summary}}: {\it{Need to model the molecule
and the contact bonding self-consistently within the same scheme, doing
justice to the molecular quantum chemistry as well as the contact surface
microstructure. The method of calculating the energy levels or the Fermi
energy is still an unresolved issue.}}

\subsection{What is the broadening due to the contacts?}

Although experimental knowledge of the contact conditions is difficult
to access, could one at least hope to model a particular idealized
contact geometry and obtain an appropriate self-energy? We obtain
the self-energy matrices $\Sigma_{1,2}$ formally by an {\it{exact}}
partitioning of the infinite metal-molecule-metal system, projecting
its single particle Green's function onto the device subspace
\cite{rDamGhosh}. $\Sigma_{1,2}$ depend on the contact surface Green's
function and the contact-molecular bondings. We obtain the couplings at
the surface by simulating a large cluster from the contact coupled to
the device and calculating its overlap and Fock matrices.  The contact
surface Green's function is calculated using a recursive Green's function
technique taking the full group theory of the FCC Au(111) crystal surface
into account \cite{rManoj,rDamle}. 

One can replace the partitioning scheme with a scattering-formalism
\cite{rPantelides,rPauls,rEmb,rJoachim} that deals with the entire
infinite system. Ideally, both methods (scattering formalism and
the NEGF prescription) should yield the same answer; however, there
is a conceptual simplification in partitioning the problem into a
``device'' part involving the electronically active molecule, and a
``contact'' part determining the lead-molecular interactions. These
involve two entirely different areas of research, quantum chemistry and
surface physics, so partitioning allows us to improve modeling each
of them independently. Moreover, NEGF naturally allows us to include
{\it{incoherent processes}}, which can be important even for a short
molecule if there are localized states that cannot be populated from
the contacts \cite{rMorkoc,rDamGhosh}.

How do we know if we have modeled our contacts correctly? The overall
``shape'' of the molecular I-V can be obtained approximately without
getting the bonding or the quantum chemistry right. One excellent
benchmark is the quantum point contact (QPC) \cite{rBrandbyge}, the I-V characteristic
of which is experimentally measured to be ohmic, with a conductance
quantized in units of $G_0 = 2e^2/\hbar \approx 77 \mu$S \cite{rQPC}.
Starting with a six atom gold chain coupled to Au(111) contacts
describing a QPC, we get a conductance quantized ohmic I-V, as expected
\cite{rDamle,rDamGhosh}. This is highly nontrivial, because
conductance quantization arises out of a molecule that is perfectly
transmitting over a band of energies between $\mu_1$ and $\mu_2$ (aside
from Fabry-Perot type oscillations).
This requires the self-energy matrices to couple the wire with the contacts
{\it{seamlessly}} without introducing any spurious reflections.
To illustrate the sensitivity of the quantization on the coupling, we
scaled the overall matrix elements by a factor of five; the resultant
I-V ceases to be ohmic, and resembles that of a resonantly conducting
system such as PDT (Fig. 5 (i)). 

\begin{figure} 
\vspace{2.1in} 
%\hskip -3cm{\vskip 4cm\special{psfile=ivs.epsi 
%\hskip -1.6cm{\special{psfile=ivs.ps
{\includegraphics{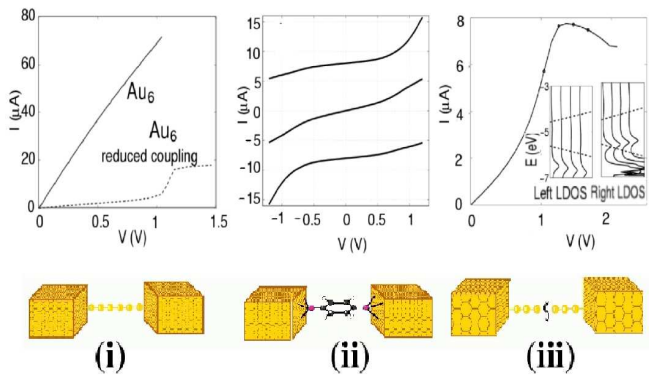}} 
\vskip 0.2cm 
\caption{Calculated two-terminal I-V characteristics for different molecular
geometries: (i) ohmic I-V with a quantized conductance (adapted from
\protect\cite{rDamle}) for a quantum point contact (QPC) consisting of
a six-atom gold chain connected to Au(111) contacts; (ii) a symmetric,
resonant I-V for PDT that turns asymmetric (upper and lower curves)
(adapted from \protect\cite{rAvik}) on altering the relative coupling
strengths to the contacts; (iii) negative differential resistance
(NDR) in the I-V for a QPC with a barrier in the middle (adapted from
\protect\cite{rDamGhosh}).} 
\label{f5} 
\end{figure}

The above exercise is a good check of the accuracy of the contact broadenings.
Once the surface Green's functions are deemed to be correct, it is an easy
matter to replace the Au$_6$ molecular cluster with the molecule of
choice and proceed with calculating its I-V.

\medskip \noindent{\bf{Summary:}}{\it{The QPC can be used as a benchmark
for testing out the self-energy matrices. The couplings at the surface and
the contact surface Green's functions need to be calculated accurately,
including the overall group theory of the metal crystal comprising
the leads}}.

\subsection{What is the spatial profile of the Laplace potential?}

The electrostatic potential profile across the molecule can be separated
into two parts: (i) the {\it{Laplace}} part describes the influence of 
the contact geometries in the absence of
charges and screening effects; (ii) the {\it{Poisson}} part involves
screening by the charges, and is determined by the charging energy of the
molecule. In this section, we will concentrate on the Laplace part, and 
address the charging-related issues associated with the Poisson part in
the next section.

The Laplace part of the potential profile is set by the relative capacitances
of the contacts, and can be determined rigorously by solving the 3-D
Laplace equation with the correct potential boundary conditions for the
contacts. The Laplace part is characterized by the voltage-division 
factor $\eta$ \cite{rDatta} which describes the proportion in which the
applied voltage drops across the various contact-molecular interfaces. 
A convenient way to analyze the potential profile across the molecule is
to incorporate the voltage-division factors in the definitions of the
contact electrochemical potentials $\mu_{1,2}$, which would keep the molecular
levels themselves fixed under drain bias in the absence of charging
(next section) and move them under gate bias alone. For a two-terminal 
device appreciable current flow requires the capacitive couplings with the source
and drain electrodes to be roughly equal (although their resistive (quantum) couplings
$\Gamma_{1,2}$ could still be quite different), leading usually to
$\eta \approx 0.5$. 

Although the Laplace part involves essentially nineteenth century
classical electrostatics, it can substantially influence device I-V
characteristics. In a gated three-terminal device, for instance, a good gate
control mechanism in a well-designed ballistic conductor essentially
involves trying to keep the charge density near the source end of the
conductor constant by pinning the device DOS to the source chemical
potential \cite{rLundstrom}. This gives an effective $\eta \ll 1$,
so that as the drain chemical potential ventures into the HLG under
bias, the current starts to saturate \cite{rFET}. Note that this saturation
mechanism is entirely different from saturation in two-terminal
molecular devices, which occurs when either contact chemical potential
has just crossed a molecular level and another level has not yet kicked
in. In contrast to the two-terminal I-V, the three-terminal
characteristic is asymmetric with respect to source-drain bias.  Since
the gate determines the position of the equilibrium Fermi energy
through the Laplace solution, it leads to gate-modulation of the ON
current in such devices.

A gate can influence the molecular electronic properties in a 
variety of ways: it could have a purely electrostatic effect on the
channel charge and correspondingly the molecular levels, as described
above. However, as pointed out by Damle {\it{et al.}} \cite{rFET},
good electrostatic gate control is not possible for a 10 $\AA$ molecule
such as PDT, unless the gate oxide is prohibitively thin. Additionally,
the gate can alter the properties of the contact-molecular interfaces
(Schottky-barrier type effects \cite{rSBFET}), or even alter the conformations of
the active molecule \cite{rRak}, all of which could affect the
shape of the molecular I-V. 

Another example where the Laplace solution itself can influence the 
molecular I-V characteristic involves conduction mechanisms that require
the alignment/misalignment of energy levels localized on different
parts of the molecule. For instance, the Aviram-Ratner mechanism \cite{rAviram}
involves a donor-bridge-acceptor system where the Laplace part of the
self-consistent potential aligns the levels at the two ends for positive
bias, and misaligns them for negative bias, leading to a strongly 
rectifying I-V characteristic. A similar example involves a quantum
point contact (QPC) with a stretched bond in the middle (Fig. 5(iii)). 
The defect disconnects the LDOS on its two sides, allowing them to
separately equilibrate with the two contacts and follow their respective
chemical potentials under source-drain bias. Within the window set by
$\mu_1$ and $\mu_2$, the Laplace solution causes the LDOS on the two
sides to slide past each other (figure inset). Since this amounts to
two transmission peaks sliding in and out of resonance, the resulting
I-V shows a weak negative differential resistance \cite{rDamGhosh,rPaulsso}.

\noindent{\bf{Summary}}: {\it{The Laplace part of the 3-D electrostatic
potential profile needs to be calculated using the boundary conditions
set by the electrodes. The Laplace potential can significantly affect
the molecular I-V characteristic, by aligning or misaligning different
parts of the molecular LDOS or by pinning the molecular DOS to one of
the electrodes in the presence of a gate terminal.}}

\subsection{What is the charging energy?}

The Poisson part of the potential profile carries information about
screening and charging inside the molecule, and is characterized by
$U_0D_0 = C_Q/C_\Sigma$, where $U_0 = e^2/C_\Sigma$ is the Coulomb
charging energy, $D_0$ is the molecular gap DOS, the quantum capacitance
$C_Q = e^2D_0$, and $C_\Sigma$ describes the total electrostatic
capacitive coupling to the various electrodes. The charging energy $U_0$
describes the ``ease'' with which the molecular levels can be filled
or emptied, and tends to drag out the I-V characteristic as shown in
Fig. 1. Note that the effect of charging can be looked upon as 
a voltage-dependent $\eta(V)$.  The net capacitance $C_\Sigma$
contributing to $U_0$ is determined by the geometry and dielectric
constant of the molecule and the electrodes. For instance, while 10
$\AA$ InAs spherical quantum dots have low charging energies $\sim 100$
meV \cite{rBanin}, a 10 $\AA$ isolated PDT molecular wire has a much
larger charging energy $\sim 3-4$ eV, making it a lot harder to cross
a level with a drain bias.  At even higher charging energies $U_0 \gg
\Gamma_{1,2}$, one can get many-body effects such as Coulomb Blockade
and Kondo resonance.  While some Coulomb-Blockade type effects can be
captured within an effective one-particle self-consistent field model
within an unrestricted calculation \cite{rCRC} (such as unrestricted
Hartree-Fock or spin density functional theory), doing justice to these
problems requires us to go beyond the one-particle prescription we have
used so far.

The Poisson solution describes the effect of adding or removing charge
from the molecule, as well as the effect of redistribution of charge
within the molecule, responsible for screening of the applied voltage.
The efficiency of the screening process depends on the amount of
material available for the reorganization of the charges. For a
molecular wire much thinner than the Debye length, the net
electrostatic potential is essentially given by the Laplace solution of
the previous section, leading to a ramp-like potential profile
\cite{rDamle,rDamGhosh,rAlbert,rNitzan,rLangAvouris}. In contrast, a
thick metallic wire allows sufficient screening, yielding a potential
profile that is essentially flat \cite{rDamGhosh,rAlbert,rNitzan}. Such
a flat potential profile can be obtained even with a thin molecular
wire if the latter is embedded in a dielectric medium as part of a
self-assembled monolayer (SAM), in which case the neighboring wires
screen the potential profile. One way to model a SAM would be disallow
any transverse variations in charge or electrostatic potential, which
would amount in effect to solving the 1-D Poisson equation. This would
give a highly screened potential profile even with a thin molecular
wire \cite{rRatnerMujica,rAlbert}.

Charging can lead to very interesting effects, such as the
creation of an asymmetric I-V with a spatially symmetric molecule
\cite{rReichert,rUrbina}. Consider a symmetric molecule with unequal
resistive (quantum) couplings to the two contacts ($\Gamma_1 \neq
\Gamma_2$). Near the onset of current conduction through a HOMO level,
a negative bias on the strong contact keeps it filled, while a positive
bias empties it. Since the molecule gets positively charged one way
but not the other, the I-V is dragged out asymmetrically such that
one gets a lower current for positive bias on the stronger contact.
Interestingly for conduction through a LUMO level, the sense of the I-V
asymmetry reverses \cite{rAvik}, because one now needs to fill the LUMO
to charge up the molecule. This allows us to identify the nature of the
conducting molecular orbital, which is important given that different
orbitals conduct quite differently. For PDT, STM data \cite{rDatta}
(the STM tip being the weaker contact) seems to indicate HOMO-based
conduction.  Our results (Fig. 5 (ii)) qualitatively match the I-V
characteristics obtained by Reichert {\it{et al.}} \cite{rReichert},
with an initially symmetric I-V that turns weakly asymmetric on drawing
either contact away from the molecule.

The total electrostatic potential profile is the sum of the Laplace
and Poisson parts, which have different effects on the molecular I-V.
For typical break-junction/STM measurements on molecules such as PDT,
$C_Q/C_\Sigma = U_0D_0 \ll 1$, and the overall potential profile is
essentially given by the Laplace potential with the net voltage dividing
up as the {\it{capacitance ratio}} $\eta$ between the electrodes. If
however the molecule is so strongly coupled to the substrate that there
are appreciable MIGS, $D_0$ becomes large enough that $C_Q/C_\Sigma$
starts to become important. In the limit of $C_Q/C_\Sigma \gg 1$, the
I-V gets dragged out substantially by charging (Fig. 1), and the final
self-consistent voltage divides between the source and drain contacts
according to their {\it{resistance ratio}} $\Gamma_2/\left(\Gamma_1 +
\Gamma_2\right)$. This situation arises for metal nanoclusters probed
by an STM tip. Since the resistance is much larger at the STM end,
much of the applied voltage drops across the STM-cluster gap, so that
the cluster levels remain pinned to the substrate, and the STM chemical
potential alternately scans the HOMO and LUMO levels under opposite drain bias.

\medskip \noindent{\bf{Summary:}} {\it{The charging energy can 
turn an otherwise symmetric I-V into an asymmetric one. Given a spatially
symmetric molecule, we predict a larger current for positive bias on
the stronger contact and HOMO conduction, while for LUMO conduction,
the sense of the I-V asymmetry is reversed.}}

\section{Computational issues} A brief discussion of computational issues
is perhaps in order. The calculation of molecular conductance requires
two steps, (a) calculating a Fock matrix given a density matrix, and
(b) calculating a density matrix from the Fock matrix. The first step
is the most computationally challenging part, involving the evaluation
of DFT-based exchange-correlation integrals which are quite numerically
complicated, especially in sophisticated basis sets involving relativistic
core pseudopotentials.  We find our own LDA calculation of the molecular
Fock matrix in a minimal basis set to be comparable in accuracy, but
not in speed, with GAUSSIAN98.  Therefore we let GAUSSIAN98 do this
part, exploiting decades of development that have gone into it.
The second step requires contact self-energy matrices which can be
calculated once and for all for realistic contact surface structures
(e.g., Au(111) contacts) using a real space recursive Green's function
technique. The contact coupling matrices can be simulated in GAUSSIAN 98
with a finite-sized cluster, with care exercised to eliminate edge effects
on the structure of the contact surface Green's function \cite{rDamGhosh}
(a suitable localized basis describing gold would sort out this problem
automatically).

The computational challenge for us is to solve the NEGF equations,
requiring us to find the number of electrons on the molecule. Such
a requirement amounts to integrating the nonequilibrium electron DOS
all the way from the bottom of the contact band to the Fermi energy.
Since the Green's functions entering the DOS expression are highly
peaked around the molecular levels, this process involves integrating
over an energy range of several hundred volts with a millivolt accuracy
for each bias point and each step of the self-consistent procedure.
We have addressed this problem in two ways:  (i) assuming a weak
(in practice, constant) energy-dependence of the self-energy matrices
(valid for Au(111) which has an energy-independent DOS near $E_F$),
one can then perform the integrals analytically \cite{rDamGhosh}; (ii)
the nonequilibrium DOS can be divided into two groups, one lying outside
the domain of $\mu_{1,2}$, and the rest inside that window. The first
part can be integrated using a contour integration scheme \cite{rGuo},
while the part between $\mu_{1,2}$ can be calculated either by brute
force grid-based integration over the finite range between $\mu_1$ and
$\mu_2$, or by reverting to the constant $\Sigma$ approximation. Although
our calculations are performed with LANL2DZ basis sets which are somewhat
delocalized, it is preferable to employ relatively localized basis sets in
order to avoid issues related to partitioning and to be consistent with
the tight-binding approach that we are using here \cite{rBrandbyge}.
Different approximations give different values of the total electron
count, thereby affecting the equilibrium Fermi energy position, which
clearly needs more attention.  However, the approximations allow us
to obtain first-principles DFT-based I-Vs for a molecule like PDT in a
few hours on a SUN workstation, taking into account the details of the
contact geometry.

\section{Issues we haven't covered}

We have seen that by appropriately modeling each experimental geometry,
we can get qualitatively and quantitatively different conductance
properties, ranging from ohmic to rectifying, switching and saturating
I-V characteristics.
We now outline three of the issues that we have ignored so far, namely,
{\it{Conformation}}, {\it{Incoherence}} and {\it{Correlation}}.

(i) {\it{Conformation.}} One of the principal advantages of a molecule is
that it is semi-flexible. This means that the conformation of a molecule
can be altered, by transferring charge or applying an external field.
Encouraging experimental indications of a conformationally mediated
I-V have been obtained, for a fullerene-based transistor \cite{rC60},
and for the redox sidegroup-specific \cite{rChen} or vibrationally
mediated \cite{rHo} NDR measurements. We are currently investigating
the role of conformational changes, in conjunction with charging and
gate electrostatics, in modulating the molecular I-V \cite{rRak} in a
three-terminal device.

(ii) {\it{Incoherence.}} For long molecules, molecular vibrations are
important as sources of incoherent or inelastic scattering, leading
to hopping or polaronic transport (see for e.g. \cite{rFisher}). Such
inelastic effects can be naturally included in the NEGF formalism,
requiring us to introduce another self-energy matrix describing the
connection of the molecule with the source of the inelastic scattering
(a phonon bath, for example) \cite{rDamGhosh,rAlbert,rDNA}. In the tunneling
regime, in particular, inelastic scattering turns out to be crucial for
transport and dissipation in long molecular wires, such as DNA.

(iii) {\it{Correlation.}} Finally, there are examples of molecular
measurements such as the Kondo effect \cite{rPasupathy}, where the physics
of current flow cannot be captured in terms of a simple one-particle
picture, and require the incorporation of many-body correlation effects
in our model. We leave these problems for future work.

We would like to thank P. Damle, D. Janes, A. Liang, S. Lodha, M. Lundstrom, M.
Paulsson, T. Rakshit, R. Reifenberger, R. Venugopal and F. Zahid for
useful discussions. This work has been supported by the US Army Research
Office (ARO) under grant number DAAD19-99-1-0198.

\end{document}